\documentclass[runningheads]{llncs}
\usepackage[T1]{fontenc}
\usepackage{graphicx}
\usepackage{booktabs}
\usepackage{amsmath, amssymb}
\usepackage{float}
\usepackage{url}
\begin{document}

\title{SharpXR: Structure-Aware Denoising for Pediatric Chest X-Rays}



\author{
Ilerioluwakiiye Abolade\inst{1} \and
Emmanuel Idoko\inst{2} \and
Solomon Odelola\inst{3} \and
Promise Omoigui\inst{4} \and
Adetola Adebanwo\inst{5} \and
Aondana Iorumbur\inst{6} \and
Udunna Anazodo\inst{7} \and
Alessandro Crimi\inst{8} \and
Raymond Confidence\inst{7}
}

\authorrunning{I. Abolade et al.}

\institute{
Federal University of Agriculture Abeokuta, Nigeria \and
University of Lagos, Nigeria \and
University of Nigeria, Nsukka \and
University of Benin, Nigeria \and
Olabisi Onabanjo University, Nigeria \and
Federal University of Technology, Minna, Nigeria \and
McGill University, Canada \and
AGH University of Krakow, Poland
}
\maketitle              
\begin{abstract}
Pediatric chest X-ray imaging is essential for early diagnosis, particularly in low-resource settings where advanced imaging modalities are often inaccessible. Low-dose protocols reduce radiation exposure in children but introduce substantial noise that can obscure critical anatomical details. Conventional denoising methods often degrade fine details, compromising diagnostic accuracy. In this paper, we present \textbf{SharpXR}, a structure-aware dual-decoder U-Net designed to denoise low-dose pediatric X-rays while preserving diagnostically relevant features. SharpXR combines a Laplacian-guided edge-preserving decoder with a learnable fusion module that adaptively balances noise suppression and structural detail retention. To address the scarcity of paired training data, we simulate realistic Poisson-Gaussian noise on the Pediatric Pneumonia Chest X-ray dataset. SharpXR outperforms state-of-the-art baselines across all evaluation metrics while maintaining computational efficiency suitable for resource-constrained settings. SharpXR-denoised images improved downstream pneumonia classification accuracy from 88.8\% to 92.5\%,  underscoring its diagnostic value in low-resource pediatric care. 

\end{abstract}

%
%
%
%
%
%

\section{Introduction}

X-ray imaging is a cornerstone of pediatric diagnostics, especially in resource-constrained settings where access to advanced modalities like CT or MRI is limited~\cite{Power2016}. It is commonly used to detect conditions such as pneumonia and bone fractures in children. Due to heightened sensitivity to ionizing radiation, pediatric imaging relies on low-dose protocols~\cite{Nagy2023,AAPM2023}. However, these protocols significantly degrade image quality, introducing noise that can obscure subtle anatomical features critical for diagnosis~\cite{Chen2017,Zhao2024}. This is particularly concerning in children, whose smaller and less ossified structures are more easily masked by noise~\cite{Arthur2000,Rajaraman2024}. The resulting diagnostic uncertainty often leads to misdiagnoses or repeat scans, exacerbating risks in low-resource environments~\cite{Rauf2007}.

Conventional denoising methods, including traditional filters and deep learning models, often struggle with the extreme noise levels and structural sensitivity of pediatric scans~\cite{Alzubaidi2021,Suzuki2017}. Most models are trained on adult or non-medical images and generalize poorly to pediatric anatomy~\cite{Rajaraman2024}. Even established architectures like REDCNN have been shown to underperform in pediatric contexts~\cite{Chen2017}. Moreover, many methods over-smooth critical structures or fail under very low signal-to-noise ratios, impairing diagnostic utility~\cite{Rajaraman2018,Luo2022}.

To address these challenges, we propose \textbf{SharpXR}, a structure-aware dual-decoder U-Net designed for denoising pediatric low-dose chest X-rays in resource-limited settings. The architecture features two decoding branches: one optimized for smooth anatomical reconstruction and another enhanced with Laplacian-based skip connections to recover fine structures. A learnable attention-based fusion module combines these outputs, allowing the model to balance structural fidelity and noise suppression adaptively. 
Our main contributions are:

\begin{itemize}
\item We introduce \textbf{SharpXR}, a dual-decoder U-Net with structure-aware attention fusion for pediatric low-dose X-ray denoising.
\item We propose a multi-scale Laplacian enhancement module to preserve anatomical boundaries without over-sharpening.
\item We simulate Poisson-Gaussian noise on a real pediatric chest X-ray dataset to reflect realistic low-dose conditions.
\item We demonstrate state-of-the-art denoising performance and improved pneumonia classification, underscoring clinical utility.
\end{itemize}

\section{Related Work}

\subsection{Low-Dose Medical Image Denoising}

Low-dose imaging reduces radiation exposure but introduces quantum noise that can obscure anatomical structures and reduce diagnostic accuracy~\cite{Chen2017,Power2016}. Traditional denoising methods like BM3D~\cite{Dabov2007} and non-local means~\cite{Buades2005} are effective against Gaussian noise but often oversmooth clinically important features~\cite{Zhao2024}.

Deep learning models have since outperformed these methods. DnCNN~\cite{Zhang2017} introduced residual learning to estimate and subtract noise directly, while REDCNN~\cite{Chen2017} applied an encoder-decoder structure with skip connections for low-dose CT denoising. ResUNet++~\cite{Jha2019} incorporated dense skip paths and attention to preserve fine structures, and Attention U-Net~\cite{Oktay2018} used gating mechanisms to suppress irrelevant features and enhance localization.

More recent transformer-based models offer improved context modeling. Hformer~\cite{Zhang2023} uses windowed attention to recover high-frequency content in noisy CT images, and SwinIR~\cite{Liang2021} blends convolution and attention for better local-global integration. However, these methods often struggle under extreme noise levels typical of pediatric scans.

\subsection{Pediatric-Specific Challenges and Solutions}

Pediatric scans are especially sensitive to noise due to smaller, evolving anatomy~\cite{Arthur2000,Power2016}. Adult-trained models often generalize poorly, missing pediatric-specific structures~\cite{Rajaraman2024}. The AAPM Task Group 273~\cite{AAPM2023} highlights this issue, calling for pediatric-specific model validation.

Public datasets like the Kermany collection~\cite{Kermany2018} support pediatric research but lack paired clean-noisy images, limiting supervised denoising. To address this, synthetic noise is modeled using Poisson-Gaussian distributions~\cite{Foi2008,Wang2021} to simulate realistic degradation.

Few models target pediatric denoising directly. Sharp U-Net~\cite{Zunair2020} added sharpening kernels to enhance edge recovery, and Luo et al.~\cite{Luo2022} applied Laplacian-based enhancements in fluoroscopy. However, these models lack adaptive fusion to balance detail preservation and noise suppression across diverse anatomical regions.

Yoon et al.~\cite{Yoon2024} improved classification using structure-aware representations via masked autoencoders, showing the importance of preserving fine detail, though their work did not address denoising directly.

Our proposed SharpXR fills this gap with a dual-decoder U-Net and learnable fusion designed for structure-preserving denoising in pediatric chest X-rays.

\section{Method}

Denoising pediatric low-dose chest X-rays poses two major challenges: (1) the absence of paired clean-noisy datasets for supervised training and (2) the need to preserve subtle anatomical details that are critical for diagnosis. Our method addresses these challenges through realistic noise modeling and a dual-decoder network tailored for structural fidelity.

\subsection{Noise Simulation}
Because publicly available pediatric datasets lack paired low-dose and standard-dose images, we simulate realistic radiographic degradation to enable supervised training. We adopt a hybrid Poisson-Gaussian noise model~\cite{Foi2008,Chen2017,Zhang2017}. Poisson noise accounts for photon-counting variability, while Gaussian noise simulates electronic and system-related disturbances. The noisy image \[
\tilde{X} = \frac{1}{\eta} \cdot \text{Poisson}(\eta X) + \mathcal{N}(0, \sigma^2),
\]
where $X$ denotes the clean image normalized to $[0, 1]$, $\eta \in [50, 300]$ controls the Poisson rate, and $\sigma \in [5, 30]$ is the standard deviation of the additive Gaussian noise. Each parameter is sampled independently per image to reflect variability in exposure and hardware characteristics across scans.

\subsection{Proposed Architecture: Dual-Decoder U-Net with Learnable Fusion}

We propose a dual-decoder U-Net architecture designed to jointly achieve effective noise reduction and preservation of anatomical detail in pediatric chest X-rays. The model is composed of a shared encoder, two task-specific decoders, and a lightweight fusion module that adaptively blends their outputs.

\paragraph{Encoder.} The encoder follows a standard U-Net backbone using double-convolution blocks with feature channels $\{64, 128, 256, 512\}$, culminating in a bottleneck of 1024 channels. It extracts multi-scale features $\{f_1, f_2, f_3, f_4\}$ from the input image.

\paragraph{Decoder for Noise Reduction.} The first decoder, $D_{\text{denoise}}$, uses conventional skip connections to reconstruct a smoothed output image:
\[
D_{\text{denoise}}: \{f_4, f_3, f_2, f_1\} \rightarrow \hat{X}_{\text{denoise}}.
\]

This pathway is designed to suppress noise while preserving large-scale structural information.

\begin{figure}[ht]
\centering
\includegraphics[width=\linewidth]{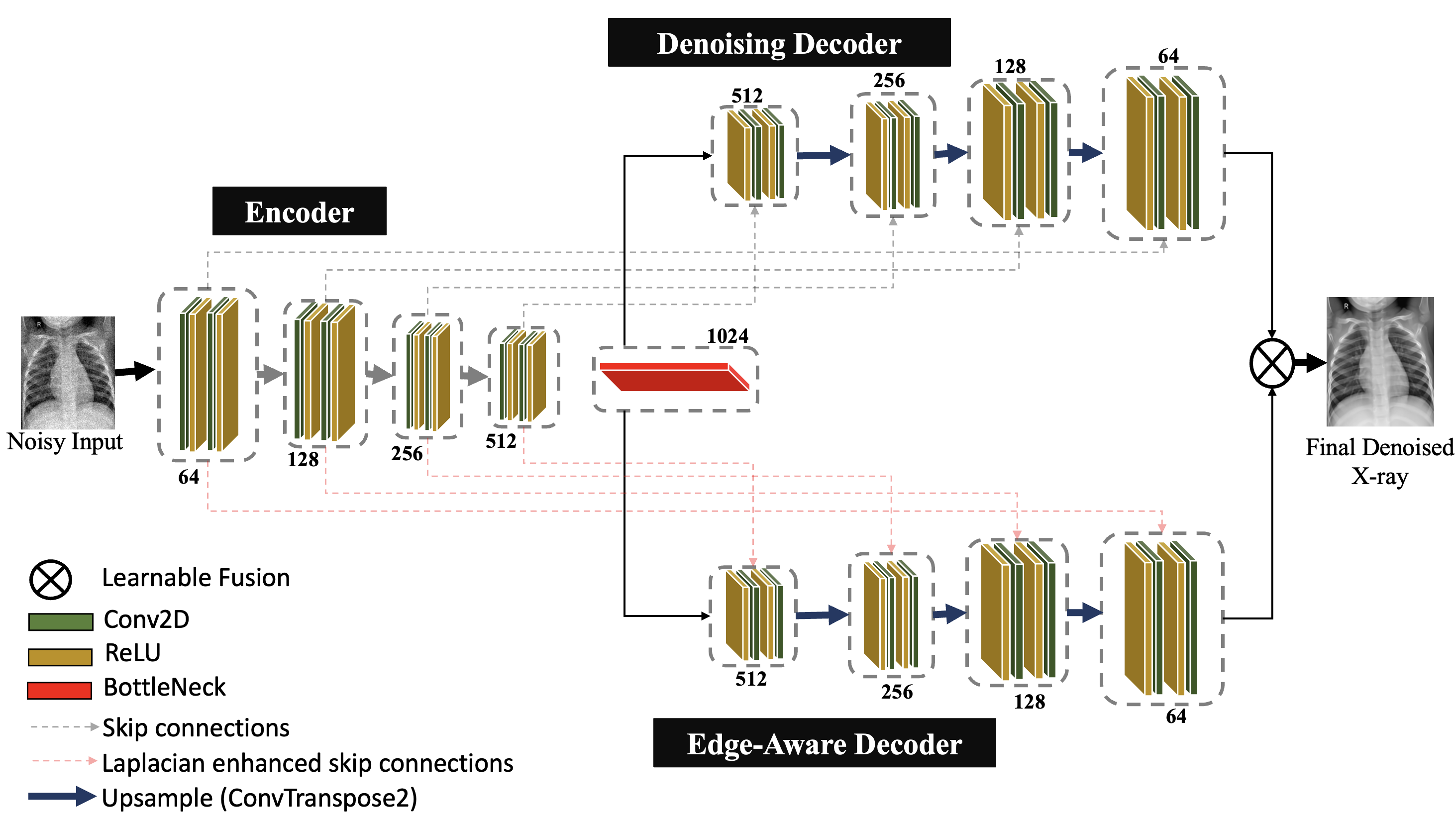} 
\caption{
\textbf{SharpXR architecture.} A dual-decoder U-Net with Laplacian-enhanced skip connections and attention-based fusion for denoising pediatric chest X-rays.
}
\label{fig:sharpxr_arch}
\end{figure}

\paragraph{Decoder for Edge Preservation.} The second decoder, $D_{\text{edge}}$, enhances the skip connections by adding Laplacian-filtered residuals, which amplify anatomical boundaries:
\[
f'_i = f_i + \mathcal{L}(f_i), \quad D_{\text{edge}}: \{f'_4, f'_3, f'_2, f'_1\} \rightarrow \hat{X}_{\text{edge}}.
\]
Here, $\mathcal{L}$ denotes the 2D Laplacian operator:
\[
\mathcal{L} =
\begin{bmatrix}
-1 & -1 & -1 \\
-1 & 8 & -1 \\
-1 & -1 & -1
\end{bmatrix}.
\]
This decoder is tailored to preserve fine anatomical features that are often degraded in low-dose acquisitions.

\paragraph{Learnable Fusion Module.} The two decoder outputs are fused using a small convolutional network that generates pixel-wise attention weights:
\[
\hat{X} = \alpha_1 \odot \hat{X}_{\text{denoise}} + \alpha_2 \odot \hat{X}_{\text{edge}},
\]
where $\alpha_1$ and $\alpha_2$ are softmax-normalized weights computed from the concatenated decoder outputs, and $\odot$ denotes element-wise multiplication. This mechanism allows the model to locally balance smoothness and edge fidelity based on contextual information.

\subsection{Loss Function and Training Strategy}

The network is optimized using the Root Mean Square Error (RMSE) loss:
\[
\mathcal{L}_{\text{RMSE}} = \sqrt{\frac{1}{N} \sum_{i=1}^{N} (\hat{X}_i - X_i)^2},
\]
which penalizes intensity deviations between the prediction $\hat{X}$ and the clean target $X$, promoting globally consistent reconstructions.

Training was performed for up to 50 epochs using the Adam optimizer (learning rate $1 \times 10^{-4}$, batch size of 4) on an NVIDIA T4 GPU. All random seeds were fixed for reproducibility.

\section{Experiments and Results}

\subsection{Dataset Description}
We utilized the Pediatric Pneumonia Chest X-ray dataset~\cite{Kermany2018}, comprising 5,856 frontal-view chest radiographs from pediatric patients aged 1--5 years. This dataset has been extensively validated in pediatric diagnostic modelling applications~\cite{Kundu2021,Yoon2024,Singh2024,AlReshan2023}. All images were resized to $256 \times 256$ pixels and normalized to $[0,1]$. Data augmentation techniques included random horizontal flipping ($p=0.5$), brightness and contrast jittering ($\pm10\%$), and random rotation ($\pm15^\circ$). The dataset was split using stratified sampling into 75\% training (4,391 images), 10\% validation (586 images), and 15\% testing (879 images).

\subsection{Evaluation Metrics}  
We evaluate SharpXR against several state-of-the-art denoising baselines, including BM3D~\cite{Dabov2007}, DnCNN~\cite{Zhang2017}, REDCNN~\cite{Chen2017}, ResUNet++~\cite{Jha2019}, Attention U-Net~\cite{Oktay2018}, Sharp U-Net~\cite{Zunair2020}, and Hformer~\cite{Zhang2023}, on the pediatric chest X-ray dataset with simulated Poisson-Gaussian noise. Performance is assessed using standard quantitative metrics: Root Mean Square Error (RMSE), Peak Signal-to-Noise Ratio (PSNR), Structural Similarity Index (SSIM), and Signal-to-Noise Ratio (SNR). Lower RMSE and higher PSNR, SSIM, and SNR values indicate better denoising performance.

\subsection{Overall Benchmark Performance}

Table~\ref{tab:pediatric_benchmark} summarizes test results. \textbf{SharpXR} achieves the best performance across all metrics.

\begin{table}[H]
\centering
\caption{Quantitative comparison of denoising models on the pediatric chest X-ray dataset. Metrics include RMSE (lower is better), PSNR, SSIM, and SNR (higher is better). The best value in each column is highlighted in bold.}
\label{tab:pediatric_benchmark}
\begin{tabular}{lcccc}
\toprule
\textbf{Model}     & \textbf{RMSE} & \textbf{PSNR} & \textbf{SSIM} & \textbf{SNR} \\
\midrule
Attention U-Net    & 0.0193 $\pm$ 0.001 & 34.44 $\pm$ 0.12 & 0.8876 $\pm$ 0.004 & 28.76 $\pm$ 0.15 \\
Sharp U-Net        & 0.0172 $\pm$ 0.001 & 35.40 $\pm$ 0.11 & 0.9261 $\pm$ 0.002 & 29.72 $\pm$ 0.14 \\
HFormer            & 0.0271 $\pm$ 0.002 & 31.46 $\pm$ 0.21 & 0.8464 $\pm$ 0.005 & 25.78 $\pm$ 0.24 \\
ResUNet++          & 0.0317 $\pm$ 0.003 & 31.89 $\pm$ 0.20 & 0.8569 $\pm$ 0.004 & 26.17 $\pm$ 0.23 \\
REDCNN             & 0.0181 $\pm$ 0.001 & 34.95 $\pm$ 0.13 & 0.9140 $\pm$ 0.003 & 29.20 $\pm$ 0.16 \\
DnCNN              & 0.0203 $\pm$ 0.001 & 33.97 $\pm$ 0.14 & 0.8945 $\pm$ 0.003 & 28.29 $\pm$ 0.18 \\
BM3D               & 0.0346 $\pm$ 0.003 & 29.45 $\pm$ 0.25 & 0.7003 $\pm$ 0.006 & 22.85 $\pm$ 0.30 \\
\textbf{SharpXR}   & \textbf{0.0170 $\pm$ 0.001} & \textbf{35.52 $\pm$ 0.12} & \textbf{0.9263 $\pm$ 0.002} & \textbf{29.84 $\pm$ 0.13} \\
\bottomrule
\end{tabular}
\vspace{-1em}
\end{table}

\subsection{Noise-Level Specific Performance}

Tables~\ref{tab:psnr_per_noise} and~\ref{tab:ssim_per_noise} present PSNR and SSIM values under various combinations of Poisson ($\eta$) and Gaussian ($\sigma$) noise. SharpXR consistently outperforms all baselines across varying noise conditions, demonstrating robustness to low and high photon counts and electronic noise.

\begin{table}[H]
\footnotesize
\centering
\caption{Denoised PSNR values per noise level.}
\label{tab:psnr_per_noise}
\begin{tabular}{lllllllll}
\toprule
$\sigma$ & $\eta$ & Attention U-Net & DnCNN & HFormer & REDCNN & ResUNet++ & Sharp U-Net & SharpXR \\
\midrule
5  & 300 & 36.367 & 36.166 & 33.151 & 37.218 & 33.777 & 37.760 & 37.819 \\
10 & 200 & 35.092 & 34.915 & 32.295 & 35.856 & 32.696 & 36.475 & 36.583 \\
15 & 150 & 33.981 & 33.812 & 31.476 & 34.736 & 31.826 & 35.445 & 35.569 \\
20 & 100 & 32.897 & 32.726 & 30.648 & 33.698 & 30.803 & 34.491 & 34.641 \\
25 & 50  & 31.517 & 31.163 & 29.535 & 32.385 & 29.357 & 33.332 & 33.514 \\
30 & 100 & 31.428 & 31.125 & 29.522 & 32.405 & 29.378 & 33.343 & 33.534 \\
\bottomrule
\end{tabular}
\vspace{-2em}
\end{table}

\begin{table}[H]
\footnotesize
\centering
\caption{Denoised SSIM values per noise level.}
\label{tab:ssim_per_noise}
\begin{tabular}{lllllllll}
\toprule
$\sigma$ & $\eta$ & Attention U-Net & DnCNN & HFormer & REDCNN & ResUNet++ & Sharp U-Net & SharpXR \\
\midrule
5  & 300 & 0.921 & 0.932 & 0.861 & 0.946 & 0.879 & 0.951 & 0.950 \\
10 & 200 & 0.904 & 0.916 & 0.832 & 0.929 & 0.852 & 0.938 & 0.938 \\
15 & 150 & 0.888 & 0.897 & 0.809 & 0.910 & 0.833 & 0.927 & 0.927 \\
20 & 100 & 0.872 & 0.874 & 0.779 & 0.891 & 0.809 & 0.914 & 0.914 \\
25 & 50  & 0.848 & 0.833 & 0.744 & 0.864 & 0.778 & 0.896 & 0.898 \\
30 & 100 & 0.846 & 0.828 & 0.742 & 0.864 & 0.776 & 0.892 & 0.897 \\
\bottomrule
\end{tabular}
\vspace{-2em}
\end{table}

\subsection{Qualitative Results}

\begin{figure}[H]
\centering
\includegraphics[width=\textwidth]{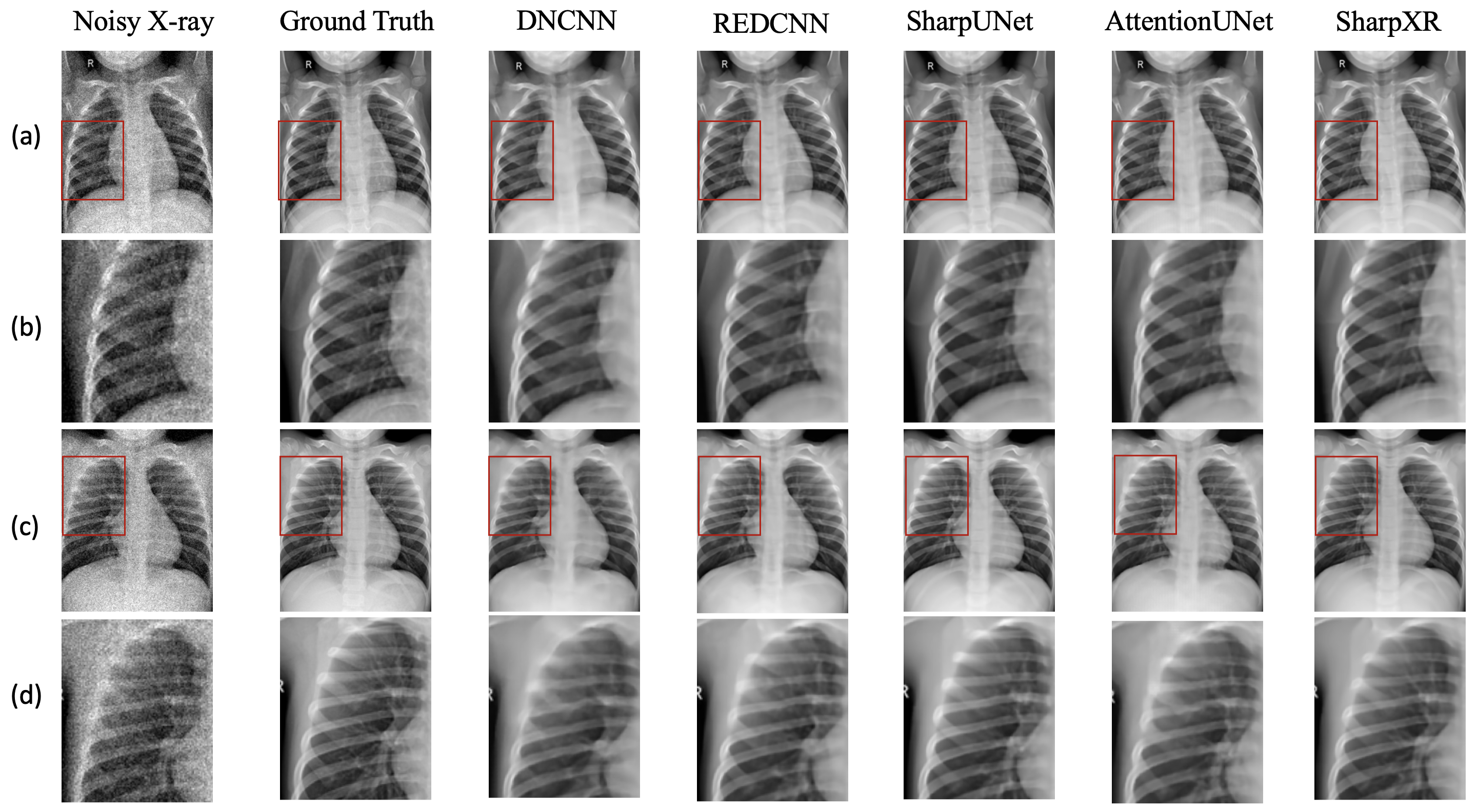}
\caption{Qualitative comparison of denoised pediatric chest X-rays. For each case (a–d), the red box marks a region of interest that is enlarged in the row below. SharpXR preserves structural detail while effectively reducing noise.}
\label{fig:qualitative}
\end{figure}

Figure~\ref{fig:qualitative} shows qualitative results on pediatric chest X-rays. For each case (a–d), the red box in the top row indicates a region of interest, which is enlarged directly below. SharpXR removes noise while preserving diagnostic structures such as ribs and lung contours.

\subsection{Downstream Classification Performance}

We evaluated SharpXR’s clinical utility using identical classifiers trained on noisy, denoised, and clean X-rays, achieving 88.8\%, 92.5\%, and 93.7\% accuracy, respectively. This confirms that SharpXR enhances diagnostic performance nearly to clean-image levels.

\subsection{Ablation Study}

We evaluate the individual contributions of the dual decoder, Laplacian enhancement, and fusion module. As shown in Table~\ref{tab:ablation_study}, the dual decoder alone improves performance over the single-decoder baseline. Adding Laplacian filtering without fusion slightly reduces SSIM, suggesting over-sharpening. Full SharpXR, combining both with learnable fusion, achieves the best overall performance.

\begin{table}[H]
\centering
\caption{Ablation study on SharpXR components. Best values are in bold.}
\label{tab:ablation_study}
\begin{tabular}{lcccc}
\toprule
Model Variant                          & RMSE   & PSNR   & SSIM   & SNR    \\
\midrule
Single Decoder                         & 0.0175 & 35.23  & 0.9257 & 29.52  \\
Dual Decoder Only                      & 0.0173 & 35.43  & 0.9257 & 29.64  \\
Dual Decoder + Laplacian (no fusion)   & 0.0175 & 35.28  & 0.9212 & 29.56  \\
\textbf{SharpXR (Full)}                & \textbf{0.0170} & \textbf{35.52} & \textbf{0.9263} & \textbf{29.84} \\
\bottomrule
\end{tabular}
\end{table}

\subsection{Discussion}
SharpXR shows strong denoising performance across metrics and noise levels. The dual-decoder with learnable fusion maintains diagnostic structures while suppressing noise. These qualities make SharpXR promising for real-world pediatric imaging, especially in low-dose and low-resource contexts. Future studies include the collection and use of real low-dose ground truth datasets with realistic noise, which are unavailable at the current stage.

\section{Conclusion}
We proposed SharpXR, a structure-aware dual-decoder U-Net tailored for denoising pediatric chest X-rays while preserving anatomical detail. By combining Laplacian-enhanced skip connections with a learnable fusion mechanism, SharpXR balances noise suppression with structural fidelity. Experimental results on simulated low-dose data demonstrate consistent improvements over traditional and learning-based baselines. However, the absence of real clinical low-dose scans in our evaluation remains a limitation. In future work, we aim to validate SharpXR on real pediatric acquisitions, explore self-supervised training to address limited labeled data, and investigate generalization to other modalities such as CT or ultrasound in resource-constrained settings.

\section*{Acknowledgements}
We thank the ML Collective community for their generous compute resources and insightful weekly feedback, which helped shape the direction of this work. We are also grateful to AFRICAI for their mentorship. This work acknowledges support from the PNRR MUR (Italian Ministry of University and Research) project PE0000013–FAIR.

\end{document}